\documentclass[aps,twocolumn,nofootinbib,prl,superscriptaddress]{revtex4}
\usepackage{amsmath,amssymb,bm}
\usepackage{graphicx}
\usepackage{epstopdf}
\usepackage{latexsym}

\begin{document}
\title{Imaging bond order near non-magnetic impurities in square lattice antiferromagnets}
\author{Ribhu K. Kaul}
\affiliation{Department of Physics, Harvard University, Cambridge MA 02138, USA}
\author{Roger G. Melko}
\affiliation{Department of Physics and Astronomy, 
University of Waterloo, Ontario, N2L3G1, Canada}
\author{Max A. Metlitski}
\affiliation{Department of Physics, Harvard University, Cambridge MA 02138, USA}
\author{Subir Sachdev}
\affiliation{Department of Physics, Harvard University, Cambridge MA 02138, USA}
\date{August 2, 2008}
\begin{abstract}
We study the textures of generalized ``charge densities'' (scalar objects invariant under time reversal), in the vicinity of non-magnetic impurities in square-lattice quantum anti-ferromagnets, by order parameter field theories. Our central finding is the structure of the ``vortex'' in the generalized density wave order parameter centered at the non-magnetic impurity. Using exact numerical data from quantum Monte Carlo simulations on an antiferromagnetic spin model, we are able to verify the results of our field theoretic study. We extend our phenomenological approach to the period-4 bond-centered density wave found in the underdoped cuprates
\end{abstract}

\maketitle

{\em Introduction:} The response of quantum many-body systems to impurities is a rich subject with important experimental consequences. At vanishingly small concentrations, the impurities behave independently and hence experimental measurements directly probe the physics of a single isolated impurity. The response of an otherwise translationally invariant quantum system to the introduction of a single quantum impurity can teach us fundamentally new things about many-body quantum physics; perhaps the most famous example is the introduction of magnetic impurities in non-magnetic metals, the so-called Kondo problem~\cite{hewson}.  Although the role of single non-magnetic impurities is innocuous in conventional metals, it has been recognized that they can have profound consequences on strongly correlated quantum magnets, because the removal of a moment from the lattice results in an uncompensated Berry phase~\cite{sbv}. Important examples of non-magnetic impurities in quantum magnets are, Zn substitution of Cu, and  La substitution of Ce, in Cu and Ce based magnetic materials, such as YBa$_2$Cu$_3$O$_{6+x}$~\cite{alloul} and CeCoIn$_5$~\cite{naka}.

Even small amounts of frustration are known to lead to enhanced fluctuations of competing order parameters in quantum magnets. One of the most important class of such competing orders are generalized {\em ``charge density''} waves~\cite{ssrmp}. We use the phrase  {\em `` charge density''} in a very general sense, to imply any order parameter that is scalar under spin rotation and even under time reversal, and hence includes, e.g., stripes~\cite{zaanen,didier,machida,inui, kivelson} and valence bond solids~\cite{rs2,vs,poilblanc,vojta}. Non-magnetic impurities couple efficiently to these fluctuation since they break translational symmetry and like the {\em ``charge density''} waves do not carry any spin. Hence the response of a magnet to non-magnetic impurities contains important information about competing orders and their quantum fluctuations.

\begin{figure}
  \includegraphics[width=2.8in]{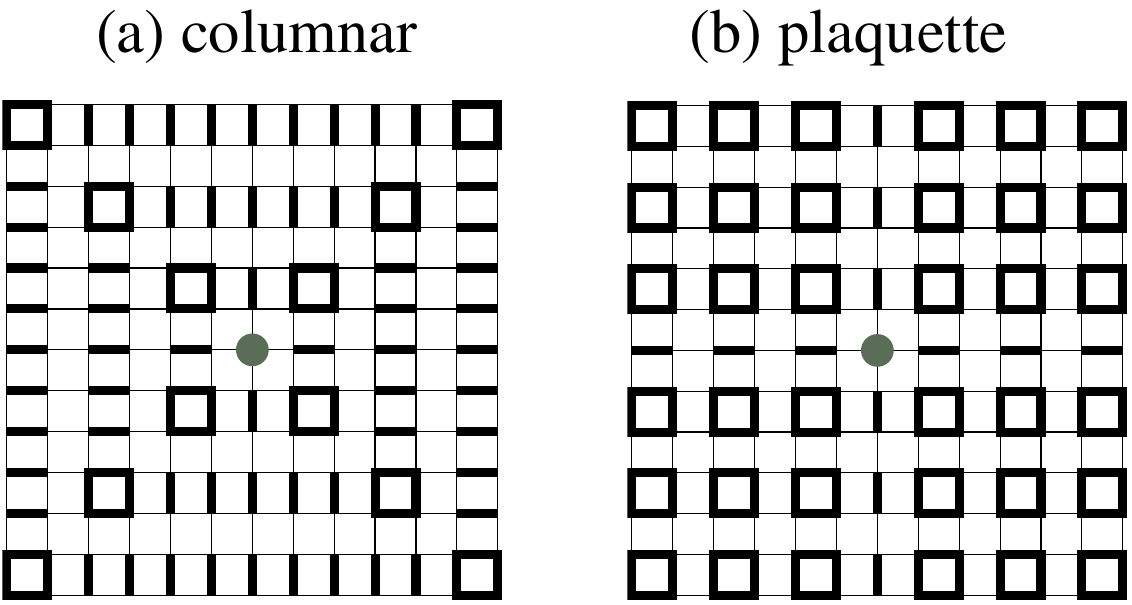}
  \caption{Cartoons of predicted modulations of VBS ``vortices'' that form around non-magnetic impurities. In both columnar and plaquette VBS phases, four domains form and are separated by domain walls of the complementary VBS pattern. Unlike the `pinwheels' depicted in earlier work \cite{levinsenthil,max2}, 
  these structures do not break any symmetry of the impurity Hamiltonian.}
  \label{fig:vortex}
\end{figure}

In this paper we address the pattern of  {\em ``charge density''} modulations in real space around a non-magnetic impurity in square lattice anti-ferromagnets through order parameter field theories as well as exact quantum Monte Carlo (QMC) simulations. One of our central results, that we verify explicitly by QMC, is the description of an impurity-centered vortex \cite{levinsenthil,max2} in the charge density order parameter as a consequence of a missing spin-1/2 moment, Fig.~\ref{fig:vortex}.  The experimental motivation for our study comes from   scanning tunnelling microscopy (STM), which can obtain detailed real space images of the described modulations in a number of materials both with and without impurities~\cite{kohsaka,hudson}. In particular, our results apply directly to Zn substitution of Cu in  square lattice anti-ferromagnets such as La$_2$CuO$_4$ and Bi$_2$Sr$_2$Dy$_{0.2}$Ca$_{0.8}$Cu$_2$O$_{8+\delta}$.

{\em Order Parameter Field Theory: }Most of our discussion will be concerned with insulating square lattice $S=1/2$ anti-ferromagnetic spin models, for which the most natural magnetic state is the collinear N\'eel state.
 Consistent with field theoretic predictions~\cite{rs2}, a fairly large body of numerical work using exact diagonlization, series expansion and quantum Monte Carlo on a variety of microscopic spin models~\cite{sandvik,melkokaul,num_vbs} has found that the competing {\em ``charge density''} wave instability of the insulating collinear N\'eel state is valence bond solid (VBS) order . The VBS order parameter is represented by
 a complex field $V ({\bf r})$, the phase containing information of the specific pattern of VBS ordering~\cite{natrev}. 
 The order parameter $V({\bf r})$ oscillates at the wavevectors $(\pi/a, 0)$ and $(0, \pi/a)$ ($a$ is the lattice spacing),
 and hence leads to  the generalized density 
\begin{equation}
\delta \rho_{V} ({\bf r}) =  \Re \bigl[ V ({\bf r}) \bigr] \sin (\pi x/a) +  \Im \bigl[ V ({\bf r}) \bigr] \sin (\pi y/a),
\label{drho2}
\end{equation}
where the origin of co-ordinates ${\bf r} = (x,y)$ is chosen at a direct lattice site.
The representation in Eq.~(\ref{drho2}) implies that the space group transformations of $V$ are as in Table~\ref{table0}, we can then write down the most general action for the fluctuations of $V$:
\begin{eqnarray}
\mathcal{S}_V &=& \int d^2 r d \tau \Bigl[ |\partial_\tau V|^2 + \widetilde{K}_1 \bigl( |\partial_x V|^2 + |\partial_y V|^2 \bigr)
 \nonumber \\ 
&+&  \widetilde{K}_2 \left((\partial_x V)^2-(\partial_y V)^2 + (\partial_x V^\dagger )^2-(\partial_y V^\dagger )^2 \right)\nonumber \\
 &+& \widetilde{s} |V|^2 + \widetilde{u} |V|^4 - \widetilde{w} (V^4 + V^{\dagger 4}) \Bigr]. \label{sv}
\end{eqnarray}
The coupling $\widetilde{w}$ chooses between columnar ($\widetilde{w} > 0$) and plaquette ($\widetilde{w}<0$) VBS ordering.

\begin{table}[!t]
\begin{tabular}{||c||c|c|c|c|c|c|c||} \hline\hline
 & $T_x$ & $T_y$&$R_{\pi/2}^{\rm dual}$ & $I_x^{\rm dual}$&$R_{\pi/2}^{\rm direct}$ & $I_x^{\rm direct}$ & $\mathcal{T}$ \\
 \hline \hline
 $V$ & $-V^\dagger$ & $V^\dagger$&$iV^\dagger$ &$V$&$iV$ & $-V^\dagger$ & $V$  \\ \hline
 $\Phi_x$ & $-i\Phi_x$ & $\Phi_x$&$\Phi_y$ & $\Phi^\dagger_x$ &$\Phi_y$ & $-i\Phi_x^\dagger$ &  $\Phi_x$  \\ \hline
 $\Phi_y$ & $\Phi_y$ & $-i\Phi_y$&$\Phi^\dagger_x$ & $\Phi_y$ &$-i\Phi^\dagger_x$ & $\Phi_y$ &  $\Phi_y$  \\ \hline
\hline
\end{tabular}
\caption{Transformation properties of the generalized charge densities. The first row is the VBS order parameter. The second and third rows are the CDW order parameters. $T_x$ : Translation  along the $x$ axis by one lattice
site; $T_y$ : Translation along the $y$ axis by one lattice
site; $R_{\pi/2}^{\rm direct}$ ($R_{\pi/2}^{\rm dual}$) : Rotation by $90^\circ$ about a direct (dual) lattice site;
$I_{x}^{\rm direct}$ ($I_{x}^{\rm dual}$) : Reflection about the $y$ axis of the direct (dual) lattice;
$\mathcal{T}$ : Time reversal.}
\label{table0}
\end{table}

In the present symmetry analysis, the effect of an impurity is modeled by the inclusion of
additional terms in the action which break the translational symmetry, but remain invariant under
$R_{\pi/2}^{\rm direct}$, $I_{x}^{\rm direct}$, and $\mathcal{T}$.
For a site centered impurity that maintains square lattice symmetry, the simplest allowed perturbation is:
\begin{displaymath}
\mathcal{S}_{{\rm imp},V} = - \lambda_1 \int d \tau \left( \frac{\partial V}{\partial x} + \frac{\partial V^\dagger}{\partial x}
+ i \frac{\partial V}{\partial y}  - i \frac{\partial V^\dagger}{\partial y} \right) \Biggr|_{{\bf r} = 0}.
\end{displaymath}
To get some intuition for the physics of $\mathcal{S}_{{\rm imp},V}$,  we make a Gaussian approximation and truncate $\mathcal{S}_V$ at quadratic order (for $\widetilde{K}_2 = 0$):
\begin{equation}
- \frac{\partial^2 V}{\partial x^2}  - \frac{\partial^2 V}{\partial y^2} + \widetilde{s} V  =  \frac{\lambda_1}{\widetilde{K}_1} \left( 
 \frac{\partial}{\partial x} - i \frac{\partial}{\partial y} \right) \delta^2 ({\bf r})
 \end{equation}
Near the impurity, this has the solution
\begin{equation}
V (|{\bf r}| \rightarrow 0) \sim \frac{e^{i \theta}}{|{\bf r}|}
\label{eq:short}
\end{equation}
showing that $\mathcal{S}_{{\rm imp},V}$ induces a vortex in the complex VBS order parameter! The divergence at ${\bf r} = 0$ will be cutoff by the discreteness of the underlying lattice. The winding of the phase of the order parameter has a direct physical implication~\cite{levinsenthil,max2}: four domains must form around the impurity separated by domain walls. The simplest pattern that does not break any symmetries present in the problem is shown in Fig.~\ref{fig:vortex} for both a columnar and plaquette VBS ordered state. 

\begin{figure}
  \includegraphics[width=3.5in]{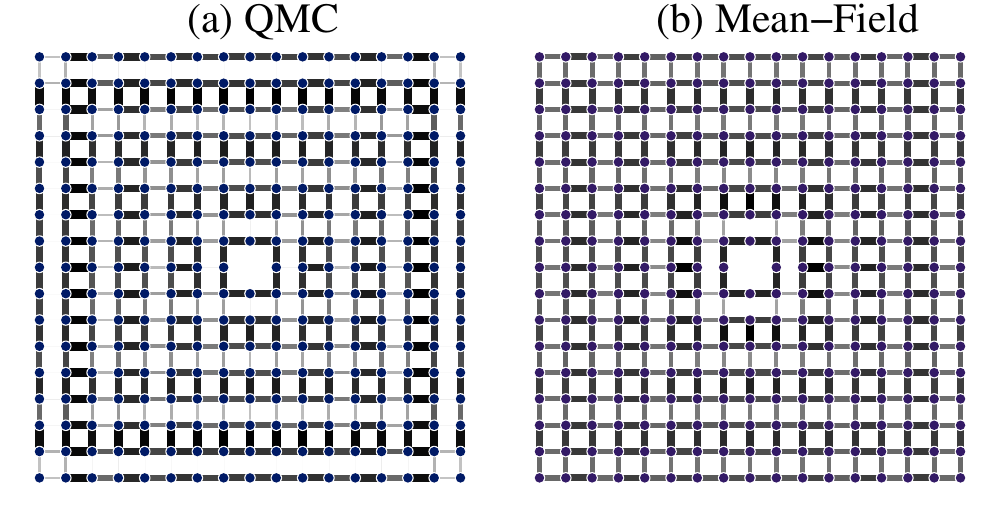}
  \caption{Comparison of exact quantum Monte Carlo data with saddle-point mean field theory. (a) Dimerization $D(x_{ij})$ of $H_{JQ}$ with a non-magnetic impurity (a missing spin) on $17\times 17$ system with open boundary conditions. $J/Q=0.1$, $T/J= 0.02$. For this value of coupling the system is in the N\'eel phase.  (b) Mean-field results for the generalized charge density $\delta\rho_V(r)$ evaluated on the bonds of a $17 \times 17$ open square lattice, evaluated from the saddle point value of $V({\bf r})$ (which contains an impurity-centered vortex) from $\mathcal{S}_V+\mathcal{S}_{{\rm imp},V}$. We have set $a=0.1$, $\widetilde{K}_1 = 1$, $\widetilde{K}_2 = 0$, $\widetilde{s}=-0.01$, $\widetilde{u}=0.3$, $\widetilde{w} = 0.1$, and the impurity coupling $\lambda_1 = 0.5$. The width and darkness of the bonds are related linearly to the plotted quantities. From Fig.~\ref{fig:32pbc} it is clear that there is a large contribution from the boundary \cite{edge}. Both the effect of the boundary and the single impurity are captured well by the mean field theory. The mean-field theory results are ``generic'' and the couplings
  were not picked to tune the agreement with Monte Carlo. }
  \label{fig:qmcmft}
\end{figure}

We have also done a numerical saddle point minimization of the full $\mathcal{S}_V + \mathcal{S}_{{\rm imp},V}$ on finite lattices and verified that for sufficient large $\lambda_1$ a vortex is indeed induced in the VBS order parameter, $V({\bf r})$, as shown in 
Fig.~\ref{fig:qmcmft}(b); note that this configuration is roughly the `negative' of the schematic in Fig.~\ref{fig:vortex}(a).

\begin{figure}[!t]
  \includegraphics[width=3.5in,trim=0 0.2in 2in 0.2in,clip=true]{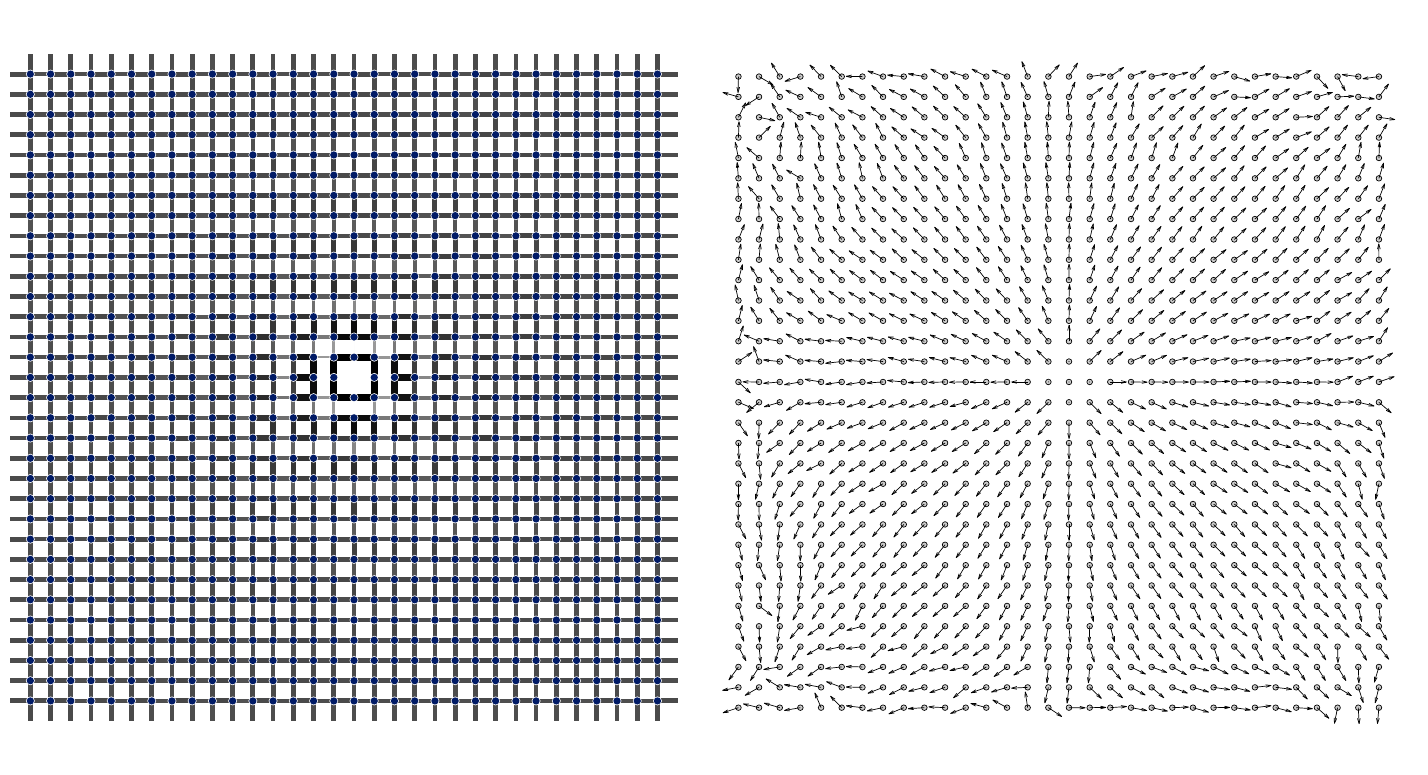}
  \caption{VBS order parameter on $32 \times 32$ system with periodic boundary conditions from QMC simulations. The figure on the left is the dimerization, $D(x_{ij})$ plotted in exactly the same way as Fig.~\ref{fig:qmcmft}. From the $D(x_{ij})$ data, we can construct a site-centered complex VBS order parameter, according to Eq.~(\ref{eq:vbsop}). The plot on the right shows the phase of this order parameter, clearly demonstrating a VBS vortex around the impurity. }
  \label{fig:32pbc}
\end{figure}

{\em Quantum Monte Carlo:} We now turn to a concrete realization of this simple, yet remarkable, effect in a microscopic model. We present the results of {\em exact} quantum Monte Carlo simulations of a model square lattice $S=1/2$ anti-ferromagnet~\cite{sandvik} with a missing spin, 
\begin{equation}
H_{JQ}= J \sum_{\langle ij \rangle}S_i\cdot S_j - Q \sum_{ijkl} \left( S_i\cdot S_j -\frac{1}{4}\right)\left( S_k\cdot S_l -\frac{1}{4}\right)\nonumber
\end{equation}
 using the stochastic series expansion method of Ref.~\onlinecite{melkokaul}. In the spin model, defining the VBS order parameter on the sites of the square lattice: 
\begin{eqnarray}
\Re[V({\bf r})] = (-1)^{{\bf r}_x}\left [ \langle S_{\bf r}\cdot S_{\bf r + x}\rangle-\langle S_{\bf r}\cdot S_{\bf r - x}\rangle \right ]\nonumber\\
\Im[V({\bf r})] = (-1)^{{\bf r}_y}\left [ \langle S_{\bf r}\cdot S_{\bf r + y}\rangle-\langle S_{\bf r}\cdot S_{\bf r - y}\rangle \right ].
\label{eq:vbsop}
\end{eqnarray}
The phase of this complex VBS order parameter contains information about the pattern of VBS ordering and it is this phase which should wind into a vortex around a non-magnetic impurity.  We shall return to the patterns in the phase shortly. First, we study the the quantity $\delta \rho_V$ on the bonds of the square lattice in our field theoretic approach. This corresponds to the dimerization $D(x_{ij}) = \langle S_i \cdot S_j\rangle$, for the quantum anti-ferromagnet. Note that it is a generalized ``charge density'' because $D(x_{ij})$ is a scalar under spin rotation and even under time reversal. We present some sample results for the dimerization patterns in Fig.~\ref{fig:qmcmft}(a) on a $17\times 17$ system with open boundary condition.  For comparison, we have also calculated $\delta \rho_V$ from our order parameter field theory by saddle point minimization of the action on $V({\bf r})$ with the same lattice geometry. Results are plotted in Fig.~\ref{fig:qmcmft}(b). It is satisfying that the results of our phenomenological theory appear in the dimerization pattern in the exact QMC simulations, including the effect of the boundary. Since the mean field solution for $V({\bf r})$ has a vortex according to Eq.~(\ref{eq:short}) and our explicit evaluation, the agreement between the mean-field theory and Monte Carlo simulations are indirect evidence of the presence of a vortex. To test the presence of the VBS vortex independently and eliminate the effects of the boundary we have simulated larger $32 \times 32$ systems with periodic boundary conditions and then explicitly constructed the complex VBS order parameter according to Eq.~(\ref{eq:vbsop}). The data in Fig.~\ref{fig:32pbc} clearly shows the winding of the phase of the VBS order parameter. Although in real materials, like the cuprates, the form of microscopic Hamiltonian is not expected to be exactly the $H_{JQ}$ model, our results are based on very general arguments and are expected to be generic to $S=1/2$ quantum anti-ferromagnets. It is also perhaps worth noting that the effects we discuss are completely quantum mechanical and are not expected to be captured in a semi-classical spin-wave approach.

{\em Period-4 Charge-density Wave:} 
Our formalism is easily extended to other density waves. As an important example we study the ubiquitous~\cite{kohsaka,hudson} density wave instability in the doped cuprates, period 4 charge density waves (CDW). We represent these
by complex order parameters $\Phi_{x}$, $\Phi_y$. These are the Fourier components of a ``{\em generalized density\/}'' modulation 
$\delta \rho ({\bf r})$. So we have
\begin{equation}
\delta \rho_{\Phi} ({\bf r}) = \mbox{Re} \bigl[ \Phi_x ({\bf r}) e^{i {\bf K}_x \cdot ({\bf r}-{\bf r}_0)} +  \Phi_y ({\bf r}) e^{i {\bf K}_y \cdot ({\bf r}-{\bf r}_0)} \bigr],
\label{drho}
\end{equation}
where ${\bf r}_0 = (a/2,a/2)$, and the wavevectors are ${\bf K}_x= (\pi/2a, 0)$, ${\bf K}_y= (0, \pi/2a)$.
Eq.~(\ref{drho}) implies
the transformation properties of $\Phi_{x,y}$ that are recorded in Table~\ref{table0}. The most general action \cite{robertson,adrian} 
in powers and gradients of $\Phi_{x,y}$ which
is consistent with the symmetries in Table~\ref{table0} is
\begin{widetext}
\begin{eqnarray}
\mathcal{S}_\Phi =  \int d^2 r d \tau  \Bigl[ |\partial_\tau \Phi_x|^2 + |\partial_\tau \Phi_y|^2 &+& 
K_1 \left( |\partial_x \Phi_x |^2 + |\partial_y \Phi_y |^2 \right )
 + K_2 \left( |\partial_y \Phi_x |^2 + |\partial_x \Phi_y |^2 \right )+
s \left( |\Phi_x|^2 + |\Phi_y|^2 \right)  \nonumber \\
&+& u \left(|\Phi_x|^2 + |\Phi_y|^2 \right)^2 + v |\Phi_x|^2 |\Phi_y|^2 - w \left (\Phi_x^4 + \Phi_y^4 + \mbox{c.c.} \right) \Bigr].
\label{sPhi}
\end{eqnarray}
\end{widetext}
The CDW ordering is stripe-like and not checkerboard for $v>0$. Also, the density modulations are bond-centered (site-centered)
for $w > 0$ ($w < 0$). In principle, linear spatial derivative terms like $\Phi_x^\ast \partial_x \Phi_x$ are also allowed,
and serve to move the ordering wavevectors away from the commensurate values ${\bf K}_{x,y}$: we will ignore such
incommensurations here, assuming the lock-in term $w$ serves to retain the commensurate value.

The CDW and VBS density waves may also couple to each other by the term,
\begin{displaymath}
\mathcal{S}_{\Phi V} = \int d^2 r d \tau \Bigl[ - \kappa V^\dagger \left( \Phi_x^2 + \Phi_x^{\dagger 2} + i \Phi_y^2
+ i \Phi_y^{\dagger 2} \right) + \mbox{c.c.} \Bigr]
\end{displaymath}

Just like we had impurity terms that coupled to $V({\bf r})$, we can write down similar terms for $\Phi_x$ and $\Phi_y$. The most relevant is a linear
terms without derivatives:
\begin{equation}
\mathcal{S}_{{\rm imp},\Phi} = - \lambda_2 \int d \tau \left( \Phi_x - i \Phi_x^\dagger + \Phi_y - i \Phi_y^\dagger \right)
\end{equation}

\begin{figure}[t!]
\centering \includegraphics[width=3.5in]{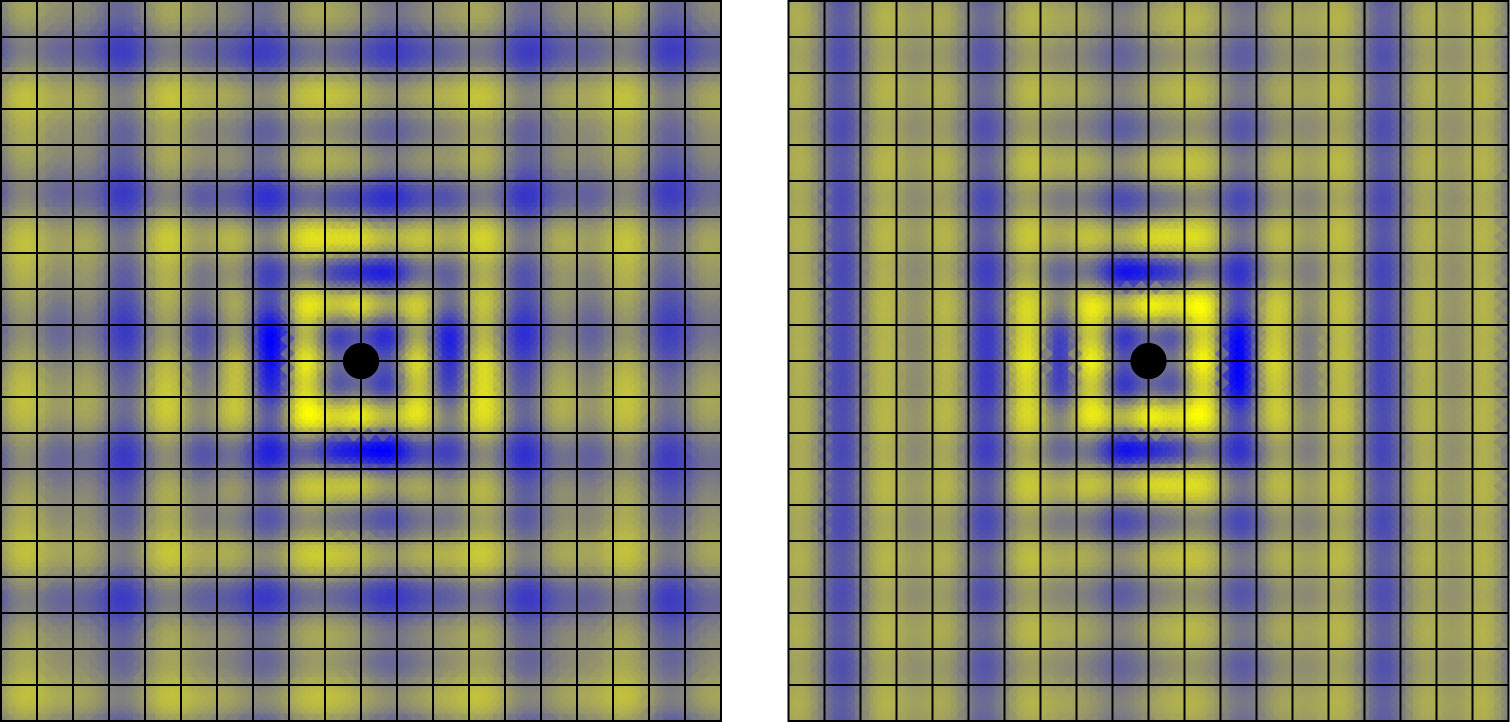}
\caption{CDW and VBS order obtained by minimizing the free energy for a $20\times 20$ lattice
with $a=0.15$; the left panel has a vortex, while the right has a domain wall in only one direction. The parameters are  $\widetilde{K}_1 = 1$, $\widetilde{K}_2 = 0$,
$\widetilde{s}=0.5 (1)$, $\widetilde{u}=1$, $\widetilde{w} = 0.1 (0.05)$, $K_1=1$, $K_2 = 1$, 
$s=-0.7 (1)$, $u=4$, $v=0.1$, $w=0$, $\kappa = 4 (8)$ and the impurity couplings $\lambda_1 = 3 (8)$ and $\lambda_2 = 0$
for the left (right) panels. 
}
\label{fig:cdw1}
\end{figure}

We can now execute numerical minimizations of the action $\mathcal{S}_\Phi + \mathcal{S}_V + \mathcal{S}_{\Phi V}
+ \mathcal{S}_{{\rm imp},V} + \mathcal{S}_{{\rm imp},\Phi}$, {\em i.e.}, we will add the period 4 CDW order parameters $\Phi_{x,y}$ to the VBS vortex configurations discussed in the discussion for the insulator. The physical idea is that at short distances around the impurity, the description of $V({\bf r})$ of the insulator
remains appropriate; for this reason, in our numerical results below, we set $\lambda_2=0$. However, at longer scales we have to account for the CDW orders, which are the primary
order parameters.
The coupling $\kappa$ will then play the role in transferring the vortex correlations
from $V$ to $\Phi_{x,y}$. Given the density wave interpretation of the VBS vortex above, we can expect corresponding
phase shifts in the period 4 density waves in  $\Phi_x$ and $\Phi_y$. 
Our numerical minimization of $\mathcal{S}_\Phi + \mathcal{S}_V + \mathcal{S}_{\Phi V}
+ \mathcal{S}_{{\rm imp},V} + \mathcal{S}_{{\rm imp},\Phi}$ led to a large number of metastable solutions,
dependent upon the initial conditions. Sample results for $\delta \rho_V+\delta \rho_\Phi$,  are shown in Fig.~\ref{fig:cdw1}; in the left panel we started from the VBS vortex and then 
ramped up the coupling to $\Phi_{x,y}$, and in the right from random initial conditions.

We have studied the textures formed by density wave order parameters around non-magnetic impurities in square lattice anti-ferromagents. We first studied insulating $S=1/2$ anti-ferromagnets for which the natural {\em ``charge density''} was described by a single complex VBS order parameter, $V({\bf r})$. We found that introducing an impurity perturbation in a phenomenological theory for $V({\bf r})$ results in the formation of a VBS vortex. Using exact QMC simulations we were able to detect this vortex explicitly. We then extended our theory to make specific predictions for the density modulations in a lightly doped anti-ferromagnet with both period-4 charge density waves $\Phi_x$ and $\Phi_y$, and  $V({\bf r})$. These results may apply close to Zn impurities in the lightly doped cuprate materials and we hope they will be tested in that case.

This research was supported by the NSF under grants DMR-0757145,  DMR-0132874 and DMR-0541988.

\end{document}